\documentclass[conference]{IEEEtran}
\usepackage{pgfplots}
\usepackage[bookmarks,colorlinks]{hyperref}
\usepackage[linesnumbered,ruled,lined]{algorithm2e}
\usepackage{enumitem}
\usepackage{algpseudocode}
\usetikzlibrary{shapes.multipart,intersections}
\usepackage{cite}
\usepackage{amsmath,amssymb,amsfonts,amsthm,steinmetz}
\usepackage{graphicx}
\usepackage{mathrsfs}  
\usepackage{textcomp}
\usepackage{acronym}
\usepackage{xcolor}
\usepackage{upgreek,xspace}

\usepackage{tikz}
\usetikzlibrary{calc}
\makeatletter
\newcommand{\gettikzxy}[3]{%
  \tikz@scan@one@point\pgfutil@firstofone#1\relax
  \edef#2{\the\pgf@x}%
  \edef#3{\the\pgf@y}%
}
\makeatother

\usepackage[draft]{todonotes}

\usepackage{esvect}

\usepackage{times}
\usepackage{bm}
\usepackage{amsmath}
\usepackage{amssymb}
\usepackage{stmaryrd}
\usepackage{babel}
\usepackage{graphics, graphicx}
\usepackage{xcolor}
\usepackage{gensymb}
\usepackage{cite}
\usepackage{enumitem}
\usepackage{url}

\usepackage{cite}

\ifCLASSINFOpdf
\else
\fi
\begin{document}
%
\title{Physics-compliant diagonal representation \\of beyond-diagonal RIS}

\author{\IEEEauthorblockN{Philipp del Hougne}
\IEEEauthorblockA{Univ Rennes, CNRS, IETR - UMR 6164 F-35000 Rennes, France\\
philipp.del-hougne@univ-rennes.fr}
}


%


\maketitle

\begin{abstract}
Physics-compliant models of RIS-parametrized channels assign a load-terminated port to each RIS element. For conventional diagonal RIS (D-RIS), each auxiliary port is terminated by its own independent and individually tunable load (i.e., independent of the other auxiliary ports). For beyond-diagonal RIS (BD-RIS), the auxiliary ports are terminated by a tunable load circuit which couples the auxiliary ports to each other.
Here, we point out that a physics-compliant model of the load circuit of a BD-RIS takes the same form as a physics-compliant model of a D-RIS-parametrized radio environment: a multi-port network  with a subset of ports terminated by individually tunable loads (independent of each other).
Consequently, we recognize that a BD-RIS-parametrized radio environment can be understood as a multi-port cascade network (i.e., the cascade of radio environment with load circuit) terminated by individually tunable loads (independent of each other). Hence, the BD-RIS problem can be mapped into the original D-RIS problem by replacing the radio environment with the cascade of radio environment and load circuit. 
The insight that BD-RIS can be physics-compliantly analyzed with the conventional D-RIS formalism implies that (i) the same optimization protocols as for D-RIS can be used for the BD-RIS case, and (ii) it is unclear if existing comparisons between BD-RIS and D-RIS are fair because for a fixed number of RIS elements, a BD-RIS has usually more tunable lumped elements.
\end{abstract}

\vskip0.5\baselineskip

\IEEEpeerreviewmaketitle

\section{Introduction}

\subsection{Background on conventional D-RIS}

The parametrization of wireless channels with reconfigurable intelligent surfaces (RISs) is at the core of the emerging paradigm shift toward smart radio environments. Conventionally, an RIS is an array of elements (oftentimes backscatter patch antennas) that each contain a tunable lumped element. Consider a scenario with $N_\mathrm{T}$ transmitting antennas, $N_\mathrm{R}$ receiving antennas and $N_\mathrm{S}$ RIS elements. The system can be described as an $N\times N$ multi-port network, where $N = N_\mathrm{A}+N_\mathrm{S}$ and $N_\mathrm{A}=N_\mathrm{T}+N_\mathrm{R}$, because we can model the tunable lumped elements as auxiliary ports terminated by tunable load impedances. The $N\times N$ multi-port network can be characterized by its scattering matrix $\mathbf{S}\in \mathbb{C}^{N \times N}$ or impedance matrix $\mathbf{Z}\in \mathbb{C}^{N \times N}$ which are related to each other via $\mathbf{Z} = Z_0 (\mathbf{I}_N + \mathbf{S}) (\mathbf{I}_N - \mathbf{S})^{-1}$, where $Z_0$ is the characteristic impedance of the single-mode transmission lines (e.g., coaxial cables) connected to the ports and $\mathbf{I}_N$ is the $N\times N$ identity matrix. The impedance matrix $\tilde{\mathbf{Z}} \in \mathbb{C}^{N_\mathrm{A}\times N_\mathrm{A}}$ that can be measured at the antenna ports is
\begin{equation}
    \tilde{\mathbf{Z}} = \mathbf{Z}_\mathcal{AA} - \mathbf{Z}_\mathcal{AS} \left(\mathbf{Z}_\mathcal{SS} + \mathbf{\Phi} \right)^{-1} \mathbf{Z}_\mathcal{SA},
    \label{eq1}
\end{equation}
where $\mathcal{A}$ and $\mathcal{S}$ denote the sets of port indices associated with antennas and RIS elements, respectively, and $\mathbf{\Phi}\in \mathbb{C}^{N_\mathrm{S}\times N_\mathrm{S}}$ is the load impedance matrix terminating the auxiliary ports~\cite{tapie2023systematic}.\footnote{The applicability of Eq.~(\ref{eq1}) to an arbitrarily complex linear passive time-invariant radio environment as well as antennas and RIS elements with arbitrary structural scattering was first noted and leveraged in Ref.~\cite{tapie2023systematic}, to the best of our knowledge. References to earlier works that were limited to free-space radio environments and/or antennas and RIS elements without structural scattering can be found in Ref.~\cite{tapie2023systematic}.} The notation $\mathbf{Z}_\mathcal{AS}$ denotes the selection of the block of $\mathbf{Z}$ whose row and column indices are the entries of the sets $\mathcal{A}$ and $\mathcal{S}$, respectively.
Conventionally, the load impedance network only connects each auxiliary RIS-element port to its own tunable load but not to the other auxiliary RIS-element ports, implying that $\mathbf{\Phi}$ is diagonal. In the following, we refer to such a conventional RIS as D-RIS (diagonal RIS).

\textit{Remark 1:} The number of parameters of the above physics-compliant model does not depend on the radio environment's complexity and there is no need to explicitly describe the radio environment or structural antenna scattering~\cite{tapie2023systematic}. All parameters can be estimated with a single full-wave simulation~\cite{tapie2023systematic}. Experimentally, the parameters can be estimated in closed-form or via gradient descent~\cite{sol2023experimentally,del2024minimal}, and are usually ambiguous (which facilitates the parameter estimation~\cite{sol2023experimentally}) unless there are at least three distinct known load impedances for each RIS element~\cite{del2024minimal}.

The wireless channel matrix $\mathbf{H} \in \mathbb{C}^{N_\mathrm{R}\times N_\mathrm{T}}$ is an off-diagonal block of the measurable scattering matrix $\tilde{\mathbf{S}} \in \mathbb{C}^{N_\mathrm{A}\times N_\mathrm{A}}$, i.e.,
\begin{equation}
    \mathbf{H} = \tilde{\mathbf{S}}_\mathcal{RT},
    \label{eq2}
\end{equation}
where $\mathcal{R}$ and $\mathcal{T}$ denote the sets of port indices associated with receiving antennas and transmitting antennas, respectively, and 
\begin{equation}
    \tilde{\mathbf{S}} = \left( \tilde{\mathbf{Z}} + Z_0 \mathbf{I}_{N_\mathrm{A}}\right)^{-1} \left( \tilde{\mathbf{Z}} - Z_0 \mathbf{I}_{N_\mathrm{A}}\right).
    \label{eq3}
\end{equation}

\textit{Remark 2:} Eqs.~(\ref{eq1}-\ref{eq3}) define the complete physics-compliant end-to-end model of a RIS-parametrized channel for an arbitrarily complex radio environment \textit{without any approximations}. Many authors make simplifying assumptions to reduce the mathematical complexity; however, throughout this paper, no simplifying assumptions will be made.

\textit{Remark 3:} Alternative physics-compliant models with lower mathematical complexity can be formulated in terms of coupled dipoles characterized by their polarizabilities~\cite{sol2023experimentally}. The number of parameters is the same as for the load-impedance-based formulation used in the present paper. Because the polarizabilities are local quantities, the polarizability-based formulation offers some unique physical insights, e.g., about the decomposition of the wireless channel into multi-bounce paths~\cite{rabault2024tacit} as well as about the effect of moving wireless entities~\cite{prod2023efficient,sol2024optimal}. Throughout the present paper we use the more widespread load-impedance-based formulation to help readers connect our insights to prior literature on BD-RIS.

\textit{Remark 4:} The theory developed in terms of impedance parameters in the present paper can be equivalently expressed in terms of scattering parameters or admittance parameters. 

\subsection{The concept of BD-RIS}

Recently, Ref.~\cite{shen2021modeling} proposed to consider a beyond-diagonal load impedance circuit for which $\mathbf{\Phi}$ is potentially a fully populated matrix. We refer to such a device as BD-RIS in this paper. Following up on Ref.~\cite{shen2021modeling}, various studies have claimed that BD-RISs outperform D-RISs, for instance, in terms of achieving more wave control with a fixed number of RIS elements~\cite{li2023reconfigurable}. However, except for Ref.~\cite{li2024beyond}, these studies were not based on physics-compliant models and even Ref.~\cite{li2024beyond} was limited to a free-space radio environment and made multiple simplifying assumptions about wave propagation. 

Theoretical papers on BD-RIS (experimental papers do not exist so far) devise new optimization algorithms for BD-RIS that essentially declare all entries of $\mathbf{\Phi}$ as optimizable parameters (up to some constraints like passivity and reciprocity). However, the optimized  $\mathbf{\Phi}$ is usually not rigorously mapped to a concrete realistic circuit that could implement the optimized  $\mathbf{\Phi}$ in practice. Thereby, the optimization is somewhat detached from the physical reality, seemingly obscuring the fundamental insights presented in the present paper.

\subsection{Contributions}

The two main theoretical insights of the present paper are as follows:
\begin{enumerate}
    \item We recognize that the BD-RIS load impedance circuit is itself a multi-port network for which a subset of ports are terminated with individually tunable loads (without connections to other ports).
    \item We recognize that a BD-RIS-parametrized wireless channel is the cascade of two multi-port networks (the radio environment and the BD-RIS load impedance circuit) terminated by individually tunable loads. An illustration of this insight is provided in the lower part of Fig.~\ref{Fig_main}. In other words, \textbf{we can map the BD-RIS problem into the conventional D-RIS problem} by replacing the radio environment in the conventional D-RIS case with the cascade of the radio environment and the load impedance circuit in the BD-RIS case. 
\end{enumerate}

\newpage
The implications of these insights are as follows:
\begin{enumerate}
    \item There is no need to develop BD-RIS-specific optimization algorithms. In fact, considering the cascade of radio environment and load circuit enforces by construction the consideration of a concrete load circuit, guaranteeing automatically that the obtained results can be mapped to a realistic circuit.
    \item It is unclear how to make a fair comparison between the performances of D-RIS and BD-RIS. Existing papers fix $N_\mathrm{S}$ but allow $N_\mathrm{C}>N_\mathrm{S}$ such that they consider a BD-RIS that has many more tunable load impedances (and hence a much larger hardware complexity) than the benchmark D-RIS. 
\end{enumerate}

\section{The multi-port network cascade \\underlying the BD-RIS concept}

Assuming that the load circuit attached to the auxiliary RIS ports is linear, \textit{irrespective of its detailed implementation} (e.g., ``fully connected'', ``group-connected'', randomly connected), it can be understood as a multi-port network with $N_2=N_\mathrm{S}+N_\mathrm{C}$ ports, where $N_\mathrm{C}$ is the number of tunable load impedances in the load circuit. Hence, the load circuit can be characterized by its impedance matrix $\mathbf{Z}^\mathrm{LC} \in \mathbb{C}^{N_2 \times N_2}$. $N_\mathrm{C}$ of the load circuit's ports are terminated with individual (i.e., \textit{not interconnected}) load impedances; the set $\mathcal{C}$ contains the corresponding port indices. The set $\bar{\mathcal{S}}$ contains the indices of the remaining $N_\mathrm{S}$ ports of the load circuit that are connected to the $N_\mathrm{S}$ ports of the radio environment whose indices are contained in the set $\mathcal{S}$ defined earlier.

\begin{figure}[h]
    \centering
    \includegraphics[width=\columnwidth]{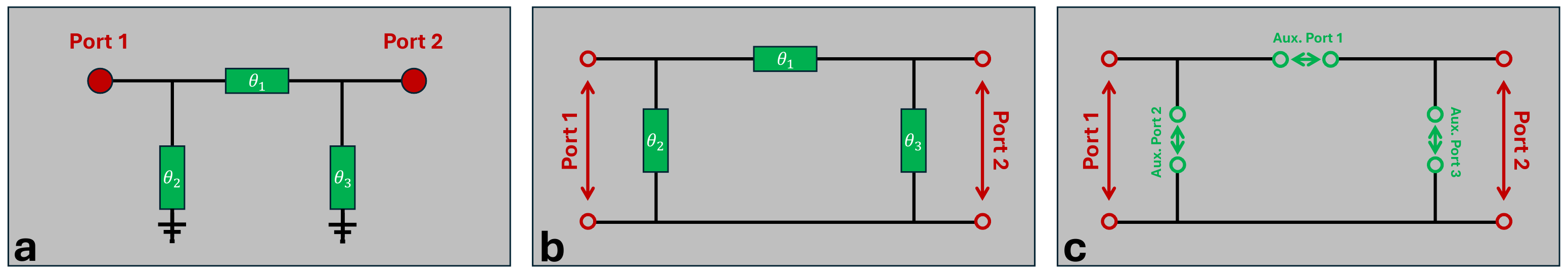}
    \caption{Clarification of the notion of a port being a ``two-terminal pair'' for a simple 2-port $\pi$-network. (a) Schematic circuit topology. (b) Detailed circuit topology clearly showing both conductors and both terminals for each port. (c) Replacement of the three impedances in (b) with three auxiliary ports that are to be terminated by individual independent load impedances. The two terminals of each port and of each auxiliary port are clearly shown. The $\pi$-network involves one series and two parallel impedances (or auxiliary load-terminated ports). }
    \label{Fig_port_explanations}
\end{figure}

\begin{figure*}
    \centering
    \includegraphics[width=1.5\columnwidth]{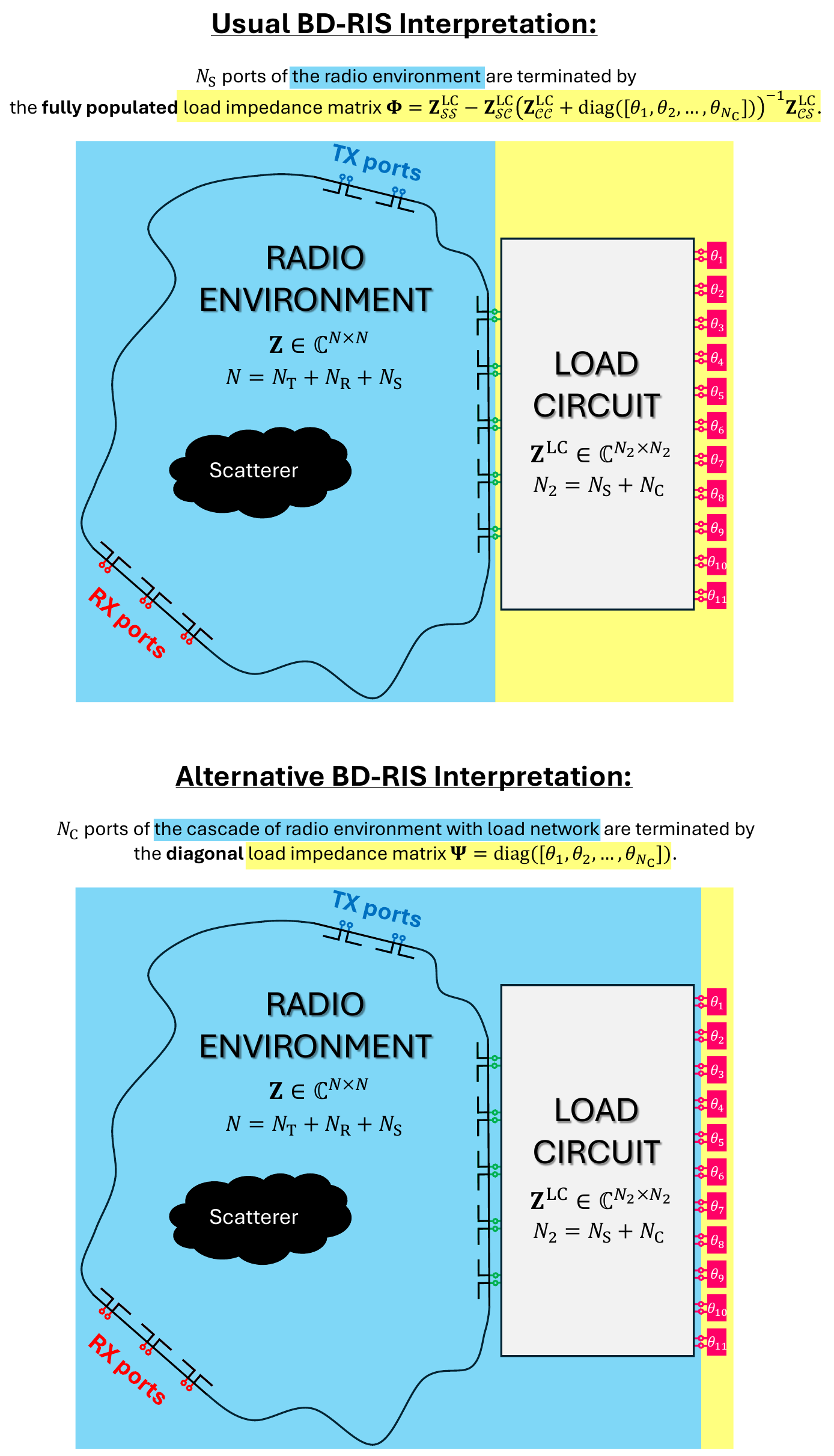}
    \caption{This figure summarizes the key insight of the present paper.   }
    \label{Fig_main}
\end{figure*}

\textit{Remark 5:} A port is defined as a ``two-terminal pair'', as highlighted in Fig.~\ref{Fig_port_explanations} and also seen in Fig.~\ref{Fig_main}, and this definition allows but does \textit{not} require that one of the two terminals of the port is grounded -- see Fig.~\ref{Fig_port_explanations}.

To start, let us determine the load impedance matrix that terminates the auxiliary RIS ports given a load circuit characterized by an impedance matrix $\mathbf{Z}^\mathrm{LC}$ as defined in the previous paragraph. This problem is analogous to that of a radio environment parametrized by a D-RIS, and hence the answer resembles Eq.~(\ref{eq1}):
\begin{equation}
    \mathbf{\Phi} = \mathbf{Z}^\mathrm{LC}_{\bar{\mathcal{S}}\bar{\mathcal{S}}} - \mathbf{Z}^\mathrm{LC}_{\bar{\mathcal{S}}\mathcal{C}} \left(\mathbf{Z}^\mathrm{LC}_\mathcal{CC} + \mathbf{\Psi} \right)^{-1} \mathbf{Z}^\mathrm{LC}_{\mathcal{C}\bar{\mathcal{S}}},
    \label{eq4}
\end{equation}
where 
\begin{equation}
    \mathbf{\Psi} = \mathrm{diag} \left( \left[ \theta_1, \theta_2, \dots, \theta_{N_\mathrm{C}} \right] \right ) \in \mathbb{C}^{N_\mathrm{C}\times N_\mathrm{C}},
\end{equation}
where $\theta_i$ is the load impedance of the $i$th load-terminated port of the load circuit. 

\textit{Remark 6:} Identifying a configuration of load impedances $\mathbf{\Psi}$ that yields a desired load impedance matrix $\mathbf{\Phi}$ is in general a non-trivial inverse-design problem (analogous to the optimization of the configuration of a D-RIS to achieve a desired property of the wireless channel).

Given $\mathbf{\Phi}$, one can insert $\mathbf{\Phi}$ into Eq.~(\ref{eq1}) and determine the physics-compliant channel matrix with Eq.~(\ref{eq2}). This corresponds to the conventional interpretation of BD-RIS-parametrized channels which explains the terminology ``beyond diagonal'': $\mathbf{\Phi}$ is not a diagonal but a ``beyond diagonal'' matrix (e.g., block diagonal or fully populated). This usual BD-RIS interpretation is summarized in the upper part of Fig.~\ref{Fig_main}. Nonetheless, to the best of our knowledge, the fact that a BD-RIS load circuit's impedance matrix $\mathbf{\Phi}$ takes the form of Eq.~(\ref{eq4}) has to date not been recognized in the literature.

Besides this usual BD-RIS interpretation, an equivalent alternative interpretation of the BD-RIS-parametrized end-to-end channel matrix exists that has, to date, not been recognized. As illustrated in the lower part of Fig.~\ref{Fig_main}, one can consider the cascade of radio environment and load circuit. This cascade is itself an $N_\mathrm{3}\times N_\mathrm{3}$ multi-port network characterized by its impedance matrix $\mathbf{Z}^\mathrm{casc} \in\mathbb{C}^{N_3 \times N_3}$, where $N_3 = N_\mathrm{A}+N_\mathrm{C}$; the ports whose indices are in the set $\mathcal{C}$ are then terminated by the diagonal load impedance matrix $\mathbf{\Psi}$. 
$\mathbf{Z}^\mathrm{casc}$ is related to $\mathbf{Z}$ and $\mathbf{Z}^\mathrm{LC}$ as follows~\cite{Jabotinski2018efficient,prod2024efficient}:
\begin{equation}
   \mathbf{Z}^\mathrm{casc} = \begin{bmatrix}  \mathbf{Z}^\mathrm{casc}_\mathcal{AA} &  \mathbf{Z}^\mathrm{casc}_\mathcal{AC}\\  \mathbf{Z}^\mathrm{casc}_\mathcal{CA} &  \mathbf{Z}^\mathrm{casc}_\mathcal{CC}   \end{bmatrix},
    \label{eq6_}
\end{equation}
where
\begin{equation}
\begin{split}
& \mathbf{Z}^\mathrm{casc}_\mathcal{AA} = \mathbf{Z}_\mathcal{AA} - \mathbf{Z}_\mathcal{AS} \mathbf{X} \mathbf{Z}_\mathcal{SA}\\
& \mathbf{Z}^\mathrm{casc}_\mathcal{AC} = \mathbf{Z}_\mathcal{AS} \mathbf{X} \mathbf{Z}^\mathrm{LC}_{\bar{\mathcal{S}}\mathcal{C}} \\
& \mathbf{Z}^\mathrm{casc}_\mathcal{CA} = \mathbf{Z}^\mathrm{LC}_{\mathcal{C}\bar{\mathcal{S}}} \mathbf{X} \mathbf{Z}_\mathcal{SA} \\
& \mathbf{Z}^\mathrm{casc}_\mathcal{CC} = \mathbf{Z}^\mathrm{LC}_\mathcal{CC} - \mathbf{Z}^\mathrm{LC}_{\mathcal{C}\bar{\mathcal{S}}} \mathbf{X} \mathbf{Z}^\mathrm{LC}_{\bar{\mathcal{S}}\mathcal{C}}\\
\end{split}
    \label{eq7_}
\end{equation}
and
\begin{equation}
    \mathbf{X} = \left( \mathbf{Z}_\mathcal{SS} + \mathbf{Z}^\mathrm{LC}_{\bar{\mathcal{S}}\bar{\mathcal{S}}} \right)^{-1}.
    \label{eq8_}
\end{equation}

\textit{Remark 7:} Equivalent expressions to Eqs.~(\ref{eq6_}-\ref{eq8_}) in terms of the corresponding scattering parameters are known as the ``Redheffer star product'' and can be found, for instance, in Refs.~\cite{redheffer_inequalities_1959,simpson_generalized_1981, chu_generalized_1986, overfelt1989alternate,prod2024efficient}.

Given $\mathbf{Z}^\mathrm{casc}$ and $\mathbf{\Psi}$, we obtain an alternative expression for the measurable impedance matrix $\tilde{\mathbf{Z}}$:
\begin{equation}
    \tilde{\mathbf{Z}} = \mathbf{Z}^\mathrm{casc}_\mathcal{AA} - \mathbf{Z}^\mathrm{casc}_\mathcal{AC} \left(\mathbf{Z}^\mathrm{casc}_\mathcal{CC} + \mathbf{\Psi} \right)^{-1} \mathbf{Z}^\mathrm{casc}_\mathcal{CA},
    \label{eq6}
\end{equation}
from which we can obtain the wireless end-to-end channel matrix $\mathbf{H}$ as before using Eqs.~(\ref{eq2}-\ref{eq3}). Importantly, recall that $\mathbf{\Psi}$ is a diagonal load impedance matrix.
The key result of the present paper is Eq.~(\ref{eq6}). Comparing Eq.~(\ref{eq1}) and Eq.~(\ref{eq6}) reveals that the BD-RIS problem can be mapped into the conventional D-RIS problem using the following analogies:
\begin{subequations}
    \begin{equation}
    \mathbf{Z} \rightarrow \mathbf{Z}^\mathrm{casc}.
    \end{equation}
        \begin{equation}
    \mathbf{\Phi} \rightarrow \mathbf{\Psi}.
    \end{equation}
    \label{eq7}
\end{subequations}
Of course, under the assumption of a trivial load circuit for which each auxiliary RIS port is terminated with an individual load impedance, the generic formulation from Eq.~(\ref{eq6}) would collapse to that of Eq.~(\ref{eq1}) because $\mathbf{Z}^\mathrm{casc}$ would simply equal $\mathbf{Z}$.

\section{Implications}

The first and most obvious implication of the insights derived in the present paper is that there is no need to develop new optimization algorithms for BD-RIS. For any realistic BD-RIS implementation, the load circuit (and hence its characterization via $\mathbf{Z}^\mathrm{LC}$) must be known such that one can always determine $\mathbf{Z}^\mathrm{casc}$ and use Eq.~(\ref{eq7}) to map the BD-RIS problem into the original D-RIS formulation.

The second implication is that the insights derived in the present paper raise questions about the fairness (or practical relevance) of existing comparisons between BD-RIS and D-RIS. Leaving aside the fact that existing comparisons are not or only partially compliant with physics, a fundamental question is whether the comparison should be for a fixed number of RIS elements $N_\mathrm{S}$ or for a fixed number of tunable load impedances $N_\mathrm{C}$. For D-RIS, $N_\mathrm{S}=N_\mathrm{C}$ whereas for the BD-RIS types considered to date, $N_\mathrm{S}<N_\mathrm{C}$. Existing comparisons are for fixed $N_\mathrm{S}$ such that a BD-RIS benefits from having drastically more tunable load impedances than a D-RIS. However, arguably the number of tunable load impedances is a limiting critical hardware aspect that is at least as important as the number of RIS elements.

\section{Conclusion}
The consideration of BD-RIS has enriched the RIS literature by generalizing the termination of the auxiliary RIS ports to arbitrarily complex tunable load circuits. However, prior to the present paper, the implications of the BD-RIS concept in terms of multi-port network theory were not fully appreciated. Here, we have shown that the BD-RIS problem constitutes a multi-port network cascade (the radio environment and load circuit are cascaded) that can always be mapped into the original D-RIS framework using Eq.~(\ref{eq7}). Our results imply that BD-RIS do not require the development of dedicated optimization algorithms and challenge the basis on which BD-RIS and D-RIS are compared in existing literature.

\section*{Acknowledgment}

The author acknowledges stimulating discussions with A.~Shaham.

\bibliographystyle{IEEEtran}



\end{document}